\documentclass[12pt]{article}
\usepackage{epsf,epsfig,amsmath}
\begin{document}

\title{Logical Inconsistencies in Proofs of the Theorem of Bell}

\author{Karl Hess$^1$ and Walter Philipp$^2$}

\date{$^1$ Beckman Institute, Department of Electrical Engineering
and Department of Physics,University of Illinois, Urbana, Il 61801
\\ $^{2}$ Beckman Institute, Department of Statistics and Department of
Mathematics, University of Illinois,Urbana, Il 61801 \\ }
%\date{\today}
\maketitle

\begin{abstract}
We discuss a class of proofs of Bell-type inequalities that are
based on tables of potential outcomes. These proofs state in
essence: if one can only imagine (or write down in a table) the
potential outcome of a hidden parameter model for EPR experiments
then a contradiction to experiment and quantum mechanics follows.
We show that these proofs do not contain hidden variables that
relate to time or, if they do, lead to logical contradictions that
render them invalid.
\end{abstract}

We have reported in previous publications \cite{hpp1},
\cite{hpp2}, \cite{hpp3} that the proofs of Bell-type inequalities
\cite{bell} come to a halt if setting dependent and time-related
hidden instrument parameters are admitted and we have concluded
that therefore the Bell inequalities may not be general enough to
directly relate to Einstein-Podolsky-Rosen (EPR)-type of
experiments \cite{eprex} because these may encompass certain time
dependencies. We summarize only some essential elements of our
reasoning and assume that the reader will be familiar with our
work, particularly \cite{hpp3}, and with our notation.

In this paper, we address additional proofs of Bell-type
inequalities including the well known types described in the book
of Peres (BP) \cite{peres} and a recent proof of Gill, Weihs,
Zeilinger and Zukowski (GWZZ) \cite{gwzz}. We show below that the
inclusion of time-related parameters invalidates these proofs too.

Before dealing with the proofs as given in BP and GWZZ, we briefly
review our general objections to derivations of Bell-type
inequalities when time-related parameters are involved. Consider
random variables $A=\pm1$ in station $S_1$ and $B=\pm1$ in station
$S_2$ that describe the potential outcome of spin measurements and
are indexed by instrument settings that are characterized by
three-dimensional unit vectors ${\bf a}, {\bf d}$ in station $S_1$
and ${\bf b}, {\bf c}$ in station $S_2$. The key assumption of
Bell \cite{bellbook} was that the random variables $A, B$ depend
only on the setting in the station and on another random variable
$\Lambda$ that characterizes the spin of particles emitted from a
common source. The possible choices of $\Lambda$ are restricted by
Bell \cite{bellbook} invoking Einstein locality arguments. We
present these arguments based on the following definitions and
remarks:

\begin{itemize}

\item[(a)] We define Einstein locality by the following postulate:
no influence can be exerted by actions in one station $S_1$ on
events in a spatially separated station $S_2$ (and vice versa)
with a speed that exceeds that of light in vacuo.

\item[(b)] We assume that the experiments correspond to the ideal
assumptions of Bell: The actual instrument settings ${\bf a}, {\bf
d}$ in station $S_1$ and ${\bf b}, {\bf c}$ in station $S_2$ are
chosen randomly (the randomness being guaranteed according to
taste by a computer, a person with free will, a quantum mechanical
measurement system or Tyche) and after the correlated pair has
been emitted from the source (delayed choice \cite{gwzz}).

\item[(c)] The conditions in Bell's mathematical model are not
necessary to fulfill these definitions and assumptions. Bell's
conditions, however, are sufficient and can be expressed as
follows: the source parameters $\Lambda$ do not depend on the
settings ${\bf a}, {\bf b}, {\bf c}, {\bf d}$ and the functions
$A, B$ only depend on the setting of their respective station and
not on that of the other. Bell further requires that $\Lambda$ has
a probability distribution $\rho$ that remains unchanged over the
whole run of experiments. For a given setting in each station,
this means that the random variables $A$ and $B$ occur with
frequencies that are related to $\rho(\Lambda)$ in both stations.
It is instructive to regard the pair $\Lambda, {\bf a}$ on which
the function $A$ depends as a new setting dependent parameter.
Then Bell's approach does contain setting dependent parameters.
However, their density is identical to that of $\Lambda$
\cite{bellbook}.

\item[(d)] Our mathematical model also suffices to obey Einstein
locality. However, we add to Bell's model setting and
time-dependent parameters e.g. $\Lambda_{{\bf a},t}^{*}$ in
station $S_1$ and $\Lambda_{{\bf b },t}^{**}$ in station $S_2$
that can, in any experimental run, have a frequency of occurrence
or density that is independent of that of $\Lambda$. This density
will depend only on the setting of the respective station (e.g.
$\bf a$ in $S_1$ and $\bf b$ in $S_2$) and not on the density of
the source parameters. The time dependence of the parameters is
necessary to fulfill certain additional requirements e.g. that we
have for the same setting $\bf a$ in both stations $A_{\bf a} =
-B_{\bf a}$ with probability one \cite{bell}. Clearly, our
mathematical model contains that of Bell as a special case and
still obeys Einstein locality. Naturally clocks in the different
stations can show correlated, even identical, times without
violating Einstein locality which only requires that no influences
are exerted with a speed faster than that of light in vacuo.

\end{itemize}

Below we discuss time-related parameters as we have previously
done \cite{hpp3}. In addition we present critiques of the BP and
GWZZ proofs according to the following three categories:

\begin{itemize}

\item[(i)] Time related parameters are excluded. This leads to the
classical proofs of Bell-type inequalities \cite{bellbook}. In our
opinion such models are not general enough to be compared with
actual EPR experiments.

\item[(ii)] The random variables $A, B$ and the joint probability
distributions are assumed to be time-dependent. Then, if we
account for the fact that only one pair of settings can be chosen
in $S_1$, $S_2$ at the time of measurement of any given correlated
pair, we show that Bell-type inequalities cannot be proven by the
known methods \cite{bellbook} \cite{peres}.

\item[(iii)] Probabilities for time-dependent potential (not
actual!) outcomes are considered for each of four different
settings at the same time for the same correlated pair. The
resulting inequalities (see Eq.(\ref{gpw1}) below) are always
valid. They also remain valid if potential violations of Einstein
locality are admitted. If, however, a transition is made from the
potential outcomes to the actual outcomes or data, then one needs
to include the fact that at a given time only one pair of settings
is possible. Hence we are back to case (ii).

\end{itemize}

We proceed now to review some of the essential features of the
proofs of Bell-type inequalities \cite{bellbook}. A key element in
all proofs is the quantity $\Delta$:
\begin{equation}
\Delta = +{A_{\bf a}}{B_{\bf c}} -{A_{\bf a}}{B_{\bf b}}-{A_{\bf
d}}{B_{\bf b}} -{A_{\bf d}}{B_{\bf c}} \label{pgr1}
\end{equation}
At this point, possible dependencies of $A, B$ on quantities other
than the settings ${\bf a}, {\bf b}, {\bf c}, {\bf d}$ are left
open. The Bell theorem states that no hidden parameters exist that
obey Einstein locality. To prove the theorem one first shows that
the following inequality holds if all elements of Eq.(\ref{pgr1})
obey Einstein locality:
\begin{equation}
|<{\Delta}>| \text{  } \leq \text{  } <|{\Delta}|> = 2
\label{pgr2}
\end{equation}
where $<...>$ indicates long-time averages and $|...|$ the
absolute value. Then one uses the fact that the long-time averages
of actually measured data that correspond to $|<{\Delta}>|$
contradict the inequality of Eq.(\ref{pgr2}) and one concludes
that the Bell theorem, stating that no local hidden variables can
describe the experiments \cite{bellbook}, is valid.

Any proof of the kind described above must permit that the
inequality of Eq.(\ref{pgr2}) be violated if parameters are
involved that do not obey Einstein locality. Because, if this
cannot be proven, then there is no possibility left to explain how
the experimental results can violate Eq.(\ref{pgr2}). We describe
below two ways how violations of the inequality can be
accomplished: Assume with Bell that the random variables $A, B$
depend in turn on the source parameter-random-variables $\Lambda$
that are independent of the settings (Einstein locality!). Assume
further that the settings are randomly chosen. Then one can
perform a thought experiment in which all the random variables
above assume values that then can be compared with experimental
data. Because of the assumption that $\Lambda$ does not depend on
settings one can always reorder the random variables of the
thought experiment in rows of four with the same value $\Lambda^i$
that $\Lambda$ has assumed for each row:
\begin{equation}
\Delta = {A_{\bf a}}(\Lambda^i){B_{\bf c}}(\Lambda^i) -{A_{\bf
a}}(\Lambda^i){B_{\bf b}}(\Lambda^i)-{A_{\bf d}}(\Lambda^i){B_{\bf
b}}(\Lambda^i) -{A_{\bf d}}(\Lambda^i){B_{\bf c}}(\Lambda^i) =
\pm2 \label{pgr3}
\end{equation}
This equation follows from the obvious fact that for any four
numbers $\xi, \eta, \zeta, \kappa = \pm1$ we have
\begin{equation}
\xi\eta -\xi\zeta - \kappa\zeta - \kappa\eta = \pm2 \label{ipgr3}
\end{equation}

The inequality of Eq.(\ref{pgr2}) is an immediate consequence of
Eq.(\ref{pgr3}). The difficulty of proving the inequality when
parameters are invoked that do not obey Einstein locality is also
easy to show. Just insert for $\Lambda^i$ a parameter that depends
on the setting of the other station. Then,
\begin{equation}
{A_{\bf a}}(\Lambda({\bf c})){B_{\bf c}}(\Lambda({\bf a}))
-{A_{\bf a}}(\Lambda({\bf b})){B_{\bf b}}(\Lambda({\bf
a}))-{A_{\bf d}}(\Lambda({\bf b})){B_{\bf b}}(\Lambda({\bf d}))
-{A_{\bf d}}(\Lambda({\bf c})){B_{\bf c}}(\Lambda({\bf d})) = ?
\label{pgr4}
\end{equation}
Now all terms with the same subscript ${\bf a}, {\bf b}, {\bf c}$
or ${\bf d}$ may be different and Eq.(\ref{pgr3}) as well as the
inequality of Eq.(\ref{pgr2}) cannot be proven. Notice that
Eq.(\ref{pgr4}) invokes spooky action in its purest form: for
example, only the settings $\bf a$ and $\bf c$ occur in the first
product and thus $A_{\bf a}$ and $B_{\bf c}$ are linked only by
the setting of the actual measurement and not by all potential
settings. If we had assumed that each of the functions $A_{\bf a},
A_{\bf d}$, $B_{\bf b}$ and $B_{\bf c}$ depend on all potentially
possible non-local parameters $\Lambda({\bf a}), \Lambda({\bf b}),
\Lambda({\bf c}), \Lambda({\bf d})$ then Eq.(\ref{pgr2}) holds
even though violations of Einstein locality are invoked. Such a
model cannot explain the experimental data.

Our point of view is, and this is the essence of our approach,
that the same effects that spatially non-local parameters have can
also be achieved by time-dependencies. Any difference or change in
setting requires a different time of measurement. This is because,
according to relativity (the finite speed of light), changes in
settings require a non-zero duration of time. If a time-dependence
is included in the functions $A, B$ we have:
\begin{equation}
\Delta = {A_{\bf a}}(t_1){B_{\bf c}}(t_1) -{A_{\bf a}}(t_2){B_{\bf
b}}(t_2)-{A_{\bf d}}(t_3){B_{\bf b}}(t_3) -{A_{\bf d}}(t_4){B_{\bf
c}}(t_4) = ? \label{pgr5}
\end{equation}
and again the proof does not go forward.

We have shown in our previous publications, that the use of
time-dependencies fulfills also all other requirements to produce
a model for EPR experiments. However, GWZZ \cite{gwzz} suggested
that time was simply irrelevant. Their proof is similar to the
well known BP proofs \cite{peres}. As we show below, neither the
reasoning in BP nor that of GWZZ renders time irrelevant and
violations of their proofs that can be achieved by violations of
Einstein locality can also be achieved by the introduction of
time-dependencies.

The starting point of BP is the following table of random
variables $A = \pm1, B = \pm1$ that describe the potential outcome
of spin measurements. At this point, we leave the possible
time-dependence of $A, B$ open and just indicate the possible time
(or time-period) of measurement at the beginning of each row. We
also show at the beginning of each row a random variable $\Lambda$
that represents the information on spin that is sent out from a
common source to the two spin analyzer stations. In the actual
experiments, only one setting per station can be chosen at a given
time. However, the table shows not the actual data but only the
random variables listed for the same time. Note also that the
table has only the form of a matrix but we do not perform any
matrix manipulations here. The right hand side of the table leaves
space for possible results that the rows of the tables add up to.

\begin{equation}
\left[ \begin{array}{cc}
      {\Lambda} & {t_1}\\
      {\Lambda} & {t_2}\\
      \vdots & \vdots\\
      {\Lambda} & {t_i}\\
      \vdots & \vdots\\
      {\Lambda} & {t_N} \end{array} \right]
\left[ \begin{array}{cccc}
      +{A_{\bf a}}{B_{\bf c}} & -{A_{\bf a}}{B_{\bf b}} & -{A_{\bf d}}{B_{\bf
      b}} & -{A_{\bf d}}{B_{\bf c}}\\
      +{A_{\bf a}}{B_{\bf c}} & -{A_{\bf a}}{B_{\bf b}} & -{A_{\bf d}}{B_{\bf
      b}} & -{A_{\bf d}}{B_{\bf c}}\\
      \vdots & \cdots & \cdots & \vdots\\
      +{A_{\bf a}}{B_{\bf c}} & -{A_{\bf a}}{B_{\bf b}} & -{A_{\bf d}}{B_{\bf
      b}} & -{A_{\bf d}}{B_{\bf c}}\\
      \vdots & \cdots & \cdots & \vdots\\
      +{A_{\bf a}}{B_{\bf c}} & -{A_{\bf a}}{B_{\bf b}} & -{A_{\bf d}}{B_{\bf
      b}} & -{A_{\bf d}}{B_{\bf c}} \end{array} \right]
 =
\left[ \begin{array}{c}
      ?\\
      ?\\
      \vdots\\
      ?\\
      \vdots\\
      ? \end{array} \right] \label{tb1}
\end{equation}

In order to form the necessary averages that lead to the spin pair
correlations one needs to sum the columns of the $AB$ products
above. To obtain Eq.(\ref{pgr2}), one needs to average the
row-sums. Note, however, that if actual outcomes would be
discussed and not just potential outcomes, then for each time only
one particular setting can be chosen and only one element of each
row contributes to the average of the measurement outcomes. This
point is essential and we will return to it below.

BP does not discuss the role of time but assumes in the
derivations, with Bell and others, that the values that the random
variable $\Lambda$ can assume do not depend on the settings.
Therefore Table~(\ref{tb1}) can be reordered and listed with the
same value that the random variable $\Lambda$ may assume indicated
for each row. As mentioned, the possibility of reordering is an
immediate consequence of Einstein locality of the source
parameters $\Lambda$ that is guaranteed by the special precautions
of delayed choice EPR-experiments as described e.g. in
\cite{gwzz}. One can then write the table for the values that the
random variables $A, B$ may assume together with the values that
the variable $\Lambda$ may assume in the following way :

\begin{equation}
\left[ \begin{array}{c}
      {\Lambda^1}\\
      {\Lambda^2}\\
      \vdots\\
      {\Lambda^i}\\
      \vdots\\
      {\Lambda^M} \end{array} \right]
\left[ \begin{array}{cccc}
      +{A_{\bf a}^1}{B_{\bf c}^1} & -{A_{\bf a}^1}{B_{\bf b}^1} & -{A_{\bf d}^1}{B_{\bf
      b}^1} & -{A_{\bf d}^1}{B_{\bf c}^1}\\
      +{A_{\bf a}^2}{B_{\bf c}^2} & -{A_{\bf a}^2}{B_{\bf b}^2} & -{A_{\bf d}^2}{B_{\bf
      b}^2} & -{A_{\bf d}^2}{B_{\bf c}^2}\\
      \vdots & \cdots & \cdots & \vdots\\
      +{A_{\bf a}^i}{B_{\bf c}^i} & -{A_{\bf a}^i}{B_{\bf b}^i} & -{A_{\bf d}^i}{B_{\bf
      b}^i} & -{A_{\bf d}^i}{B_{\bf c}^i}\\
      \vdots & \cdots & \cdots & \vdots\\
      +{A_{\bf a}^M}{B_{\bf c}^M} & -{A_{\bf a}^M}{B_{\bf b}^M} & -{A_{\bf d}^M}{B_{\bf
      b}^M} & -{A_{\bf d}^M}{B_{\bf c}^M} \end{array} \right]
 =
\left[ \begin{array}{c}
      \pm2\\
      \pm2\\
      \vdots\\
      \pm2\\
      \vdots\\
      \pm2 \end{array} \right] \label{tb2}
\end{equation}

By Eq.(\ref{pgr3}), each row of $AB$ products added up (with the
sign as given) equals $\pm2$. Averaging all the columns and rows
one obtains a Bell-type inequality of the type of Eq.(\ref{pgr2})
for the potential outcomes (for the values the random variables
may assume). Because all these potential outcomes may indeed
correspond to real outcomes, one may be tempted to compare the
result with EPR experiments. However, the parameter space is not
general enough to compare its result with an experiment that may
include time-dependencies. Had BP permitted time-dependencies of
the random variables and included a time index, then the elements
of the reordered rows of four in Table~(\ref{tb2}) have all
different time indices and may all assume different values. Then
the equality to $\pm2$ cannot be deduced in any row. Consequently
the proof of the inequality fails.

At this point, however, one may ask the question whether the above
arguments involving the potential values of random variables
cannot be extended to include time. Indeed they can. Write each
row with the same time index and create the whole table of
possible outcomes for a given value $\Lambda^i$:

\begin{equation}
\left[ \begin{array}{c}
      {t_1}\\
      {t_2}\\
      \vdots\\
      {t_i}\\
      \vdots\\
      {t_N} \end{array} \right]
\left[ \begin{array}{cccc}
      +{A_{\bf a}^1}{B_{\bf c}^1} & -{A_{\bf a}^1}{B_{\bf b}^1} & -{A_{\bf d}^1}{B_{\bf
      b}^1} & -{A_{\bf d}^1}{B_{\bf c}^1}\\
      +{A_{\bf a}^2}{B_{\bf c}^2} & -{A_{\bf a}^2}{B_{\bf b}^2} & -{A_{\bf d}^2}{B_{\bf
      b}^2} & -{A_{\bf d}^2}{B_{\bf c}^2}\\
      \vdots & \cdots & \cdots & \vdots\\
      +{A_{\bf a}^i}{B_{\bf c}^i} & -{A_{\bf a}^i}{B_{\bf b}^i} & -{A_{\bf d}^i}{B_{\bf
      b}^i} & -{A_{\bf d}^i}{B_{\bf c}^i}\\
      \vdots & \cdots & \cdots & \vdots\\
      +{A_{\bf a}^N}{B_{\bf c}^N} & -{A_{\bf a}^N}{B_{\bf b}^N} & -{A_{\bf d}^N}{B_{\bf
      b}^N} & -{A_{\bf d}^N}{B_{\bf c}^N} \end{array} \right]
 =
\left[ \begin{array}{c}
      \pm2\\
      \pm2\\
      \vdots\\
      \pm2\\
      \vdots\\
      \pm2 \end{array} \right] \label{tb3}
\end{equation}

Now repeat the table for all possible $\Lambda^i$ and the
Bell-type inequalities for all these possible outcomes are
confirmed. However, the number of elements of this theory, i.e. of
the above listed values that the random variables may assume,
exceeds the number of values of actual experiments by at least a
factor of four, since only one setting per station can be set at a
time in real experiments as opposed to thought experiments.
Therefore the model for the thought experiment has now no relation
to the real experiment because there are only two possibilities:
(i) if time is excluded, the derived Bell-type inequalities are
not general enough and (ii) if time is included, the derivation of
the inequalities makes it necessary to add up many more elements
than an actual experiment possibly can have. In passing we note
that there exist experimental variations with three measurements
which are subject to a slightly modified but overall similar
criticism \cite{ghsz}. Four measurements performed at the same
time (one shot experiments) that result in a contradiction are not
possible. This fact has been discussed in detail by Peres
\cite{peres}.

We proceed now to the proof of GWZZ \cite{gwzz}. This proof is a
slight modification from the one above and it shows an interesting
twist. Instead of considering the $AB$ products of the above
tables of random variables (or potential outcomes), GWZZ consider
the probabilities that the potential outcomes for $A$ and $B$ are
equal. Defining, for example, $P\{{A_{\bf a}} = {B_{\bf c}}\}$ as
the probability that ${A_{\bf a}} = {B_{\bf c}}$ they obtain
(their Eq.(6) \cite{gwzz}):
\begin{equation}
P\{{A_{\bf a}} = {B_{\bf c}}\} - P\{{A_{\bf a}} ={B_{\bf b}}\}-
P\{{A_{\bf d}} = {B_{\bf b}}\} - P\{{A_{\bf d}} = {B_{\bf c}}\}
\leq 0 \label{gr1}
\end{equation}
It is of no concern here which values the random variable
$\Lambda$ may actually assume. These inequalities are indeed
always true for the above probabilities of the lists of potential
outcomes. Simply notice that for any four random variables $\xi,
\eta, \zeta, \kappa = \pm1$ we have
\begin{equation}
P(\xi=\eta)-P(\xi=\zeta)-P(\kappa=\zeta)-P(\kappa=\eta) \leq 0
\label{gpw1}
\end{equation}
GWZZ do not introduce time into this equation and claim that time
is irrelevant for these probabilities that are related to the
potential outcomes for the same correlated pair. They claim ``we
did not compare actual outcomes under (sic) different settings at
different times, but potential outcomes under (sic) different
settings at the same time". It is indeed true that they could have
added an equal time index e.g. $t_i$ (symbolizing a sequence of
measurements at times $t_1, t_2,...,t_N$) to all the probabilities
of Eq.(\ref{gr1}) and the equation would be equally valid. The
problem of this line of reasoning, however, can be seen from the
following  fact. Because Eq.(\ref{gr1}) applies for all $\Lambda$,
it is also valid if we insert $\Lambda$'s that are a function of
all settings e.g. in the first term of Eq.(\ref{gr1}) ${A_{\bf
a}(\Lambda({\bf a}), \Lambda({\bf b}), \Lambda({\bf c}),
\Lambda({\bf d}))}{B_{\bf c}(\Lambda({\bf a}), \Lambda({\bf b}),
\Lambda({\bf c}), \Lambda({\bf d}))}$ and the same for the three
other terms. This means that Eq.(\ref{gr1}) is still valid if $A,
B$ depend on all potential non-local parameters. Therefore, the
inequality of Eq.(\ref{gr1}) cannot even be violated if all
potential nonlocal parameters are included. However, in order to
explain the experiments there needs to be a way to violate the
inequality of Eq.(\ref{gr1}). This can only be accomplished in a
logical way by making a transition from potential to actual
outcomes. One can then introduce action at a distance declaring
exactly the two settings of the actual measurement as the cause
for this action at a distance. Thus one has e.g. in the first term
of Eq.(\ref{gr1}) the functions ${A_{\bf a}(\Lambda({\bf
c}))}={B_{\bf c}(\Lambda({\bf a}))}$ and similar for the other
terms. Then all terms are different and the inequality is invalid.
Our main point is that exactly the same effect can be accomplished
by admitting time dependencies and again considering the actual
outcomes.

Denote with GWZZ the actual outcome of a given experiment by $X$
in station $S_1$ and by $Y$ in station $S_2$ and denote the
conditional probability that $X = Y$ for given settings ${\bf a},
{\bf b}$ by $P\{X=Y|{\bf a}{\bf b}\}$. Then GWZZ transform
Eq.(\ref{gr1})which is their Eq.(6) into their Eq.(7):
\begin{equation}
P\{X=Y|{\bf a}{\bf c}\} - P\{X=Y|{\bf a}{\bf b}\} - P\{X=Y|{\bf
d}{\bf b}\} - P\{X=Y|{\bf d}{\bf c}\} \leq 0 \label{gr2}
\end{equation}
If we now add an equal time index $t_i$ symbolizing a sequence of
measurements at times $t_1, t_2,...,t_N$ as GWZZ claim they can do
with impunity we have:
\begin{equation}
P_{t_i}\{X=Y|{\bf a}{\bf c}\} - P_{t_i}\{X=Y|{\bf a}{\bf b}\} -
P_{t_i}\{X=Y|{\bf d}{\bf b}\} - P_{t_i}\{X=Y|{\bf d}{\bf c}\} \leq
0 \label{gr3}
\end{equation}
However, this equation contains now a contradiction. For any given
time sequence, one can have one given setting only and not four
different given settings. As mentioned above, it is also well
known that there exists no possibility to obtain a contradiction
between a theoretical inequality and a ``one shot" experiment
\cite{peres}. Therefore, Eq.(\ref{gr3}) cannot be considered a
consequence of $N$ ``one shot" experiments. In addition, no
experiment that measures $4N$ events with $4$ different settings
all measured at exactly the same time has been performed or can
possibly be performed.

There is also a mathematical inconsistency in the derivations of
GWZZ that becomes clear from their longer paper \cite{gwzz} on
which their comment is based. GWZZ start from the equation
\begin{equation}
1\{{A_{\bf a}} = {B_{\bf c}}\} - 1\{{A_{\bf a}} ={B_{\bf b}}\}-
1\{{A_{\bf d}} = {B_{\bf b}}\} - 1\{{A_{\bf d}} = {B_{\bf c}}\}
\leq 0 \label{gr4}
\end{equation}
where they use the indicator function $1\{...\}$ for the events.
Subsequently they take expectation values on both sides to arrive
at Eq.(\ref{gr1}) which involves probabilities. However, it is the
random events and not their probabilities that are observed (in
the sense of statistics). Therefore the arguments of GWZZ should
have been carried out with all the probabilities $P$ replaced by
the indicator function $1\{...\}$. As a consequence of such a
notational change, it becomes now absolutely clear that only one
term can be taken into account in Eq.(\ref{gr2}) because only one
pair of settings can be chosen in $S_1, S_2$ at the time of
measurement of any given correlated pair. The inequality can
therefore not be derived. An assertion of the type that any $4$
probabilities can be added together is, in the present context,
meaningless from the vantage point of elementary statistics.

If time is of the essence and Eq.(\ref{gr4}) is labelled by a
given time or time sequence, then the theory for actual outcomes
(not just potential outcomes) must not add up elements that cannot
possibly correspond to the experiment. This is exactly the
situation that we have discussed in connection with the BP proofs.
A theory that (in order to form averages) adds up more elements
than the experiments can possibly contain cannot serve as a model
for these experiments.

We summarize as follows. Any theory that may be compared to EPR
experiments needs to be able to violate Bell-type inequalities
depending on some fact, be that a violation of Einstein locality,
the existence of time-dependencies or both. We have shown that
both ways are possible by using models that are based on
elementary probability theory. Time dependencies and violations of
Einstein locality have very similar consequences. Whether or not a
decision between the two can be made, in other words, whether any
of these random variables actually do exist in nature cannot
currently be decided with certainty. However, the Bell
inequalities can also not be used to decide with any certainty
against hidden parameters that are local in the sense of Einstein.

Acknowledgement: We thank Anthony Leggett and Michael Weissman for
numerous very helpful discussions and Salvador Barraza-Lopez for
carefully reading the manuscript. Our work was supported by the
Office of Naval Research N00014-98-1-0604.

\end{document}